\documentclass[preprint, twocolumn]{aastex631}

\newcommand{\photu}{photon units}

\newcommand{\galex}{{\it GALEX}}

\newcommand{\voyager}{{\it Voyager}}
\newcommand{\lya}{Ly$\alpha$}

\newcommand{\ebv}{E(B~-~V)}

\newcommand{\asec}{\hbox to 1pt{}\rlap{$^{\prime\prime}$}.\hbox to 2pt{}}
\newcommand{\amin}{\hbox to 1pt{}\rlap{$^{\prime}$}.\hbox to 1pt{}}
\newcommand{\adeg}{\hbox to 1pt{}\rlap{$^{\circ}$}.\hbox to 2pt{}}

\newcommand {\HI}        {\ion{H}{1}}   
\newcommand{\Htwo}     {H$_2$}        

\newcommand {\HeI}     {\ion{He}{1}}   

\newcommand {\GALEX}    {{\it GALEX}}

\newcommand {\etal}   {et~al.}

\begin{document}

\title{Excess Ultraviolet Emission at High Galactic Latitudes: A New Horizons View}

\author[0000-0003-4034-5137]{Jayant Murthy}
\affiliation{Indian Institute of Astrophysics, Bengaluru 560 034, India}

\author{J. Michael Shull}
\affiliation{Department of Astrophysical \& Planetary Sciences, CASA, University of Colorado, Boulder, CO 80309, USA}
\affiliation{Department of Physics \& Astronomy, University of North Carolina, Chapel Hill, NC 27599, USA}

\author{Marc Postman}
\affiliation{Space Telescope Science Institute, 3700 San Martin Drive, Baltimore, MD 21218, USA}

\author{Joel Wm. Parker}
\affiliation{Department of Space Studies, Southwest Research Institute, 1301 Walnut Street, Suite 300, Boulder, CO 80302, USA}

\author{Seth Redfield}
\affiliation{Astronomy Department and Van Vleck Observatory, Wesleyan University, Middletown, CT 06459, USA}

\author{Nathaniel Cunningham}
\affiliation{Nebraska Wesleyan University, Lincoln, NE, USA}

\author{G. Randall Gladstone}
\affiliation{Southwest Research Institute, San Antonio, TX 78238}
\affiliation{University of Texas at San Antonio, San Antonio, TX 78249, USA}

\author{Jon P. Pineau}
\affiliation{Stellar Solutions, Aurora, CO, 80011, USA}

\author{Pontus Brandt}
\affiliation{The Johns Hopkins University Applied Physics Laboratory, Laurel, MD 20723-6099, USA}

\author{Anne J. Verbiscer}
\affiliation{Department of Astronomy, University of Virginia, Charlottesville, VA 22904, USA}

\author{Kelsi N. Singer}
\affiliation{Department of Space Studies, Southwest Research Institute, 1301 Walnut Street, Suite 300, Boulder, CO 80302, USA}

\author{Harold A. Weaver}
\affiliation{The Johns Hopkins University Applied Physics Laboratory, Laurel, MD 20723-6099, USA}

\author{Richard C. Henry}
\affiliation{Johns Hopkins University, Dept. of Physics and Astronomy, Baltimore, 21218, USA}

\author{S. Alan Stern}
\affiliation{Southwest Research Institute, Space Sector, 1301 Walnut Street, Suite 300, Boulder, CO 80302, USA}

\begin{abstract}

We present new observations of the cosmic ultraviolet background (CUVB) at high Galactic latitudes ($|b| > 40^{\circ}$), made using the Alice UV spectrograph on board the New Horizons spacecraft. These observations were taken at about 57 AU from the Sun, outside much of the foreground emission affecting previous missions, and allowed a new determination of the spectrum of the CUVB between 912 -- 1100~\AA\ and 1400 -- 1800~\AA. We found a linear correlation between the CUVB and the Planck E(B~-~V) with offsets at zero-reddening of $221 \pm 11$ \photu\ at 1000~\AA\ and $264 \pm 24$ \photu\ at 1500~\AA\ ($4.4 \pm 0.2$ nW m$^{-2}$ sr$^{-1}$ at 1000~\AA\ and $5.3 \pm 0.5$ nW m$^{-2}$ sr$^{-1}$ at 1500~\AA). The former is the first firm detection of the offset in the range 912 -- 1100 \AA\ while the latter result confirms previous results from \galex, showing that there is little emission from the Solar System from 1400 -- 1800 \AA. About half of the offset may be explained by known sources (the integrated light of unresolved galaxies, unresolved stars, emission from ionized gas, and two-photon emission from warm hydrogen in the halo) with the source of the remaining emission as yet unidentified. There is no detectable emission below the Lyman limit with an upper limit of $3.2 \pm 3.0$ \photu.

\end{abstract}


\section{Introduction} \label{sec:intro}

Measurements of electromagnetic radiation backgrounds are historically important in astrophysics 
as constraints on the amounts of thermal energy in the microwave background, hot gas in the 
interstellar medium (ISM), and radiant energy produced by stars, galaxies, and black holes 
throughout cosmic history.  The cosmic microwave background radiation (CMB) discovered in 
1965  \citep{Penzias1965, Dicke1965} solidified the ``hot big bang" model and led to 
many future applications in precision cosmology \citep{Peebles2020}.  Observations of thermal 
anisotropies in the CMB are critical for deriving fundamental cosmological parameters from 
the {\it Wilkinson Microwave Anisotropy Probe} \citep{Bennett2003, Spergel2003} and 
the {\it Planck} satellite \citep{PlanckCollaboration2014, PCV2020}.   

An isotropic X-ray (2-10 keV) background, first detected in rocket observations \citep{Giacconi1962}, was initially interpreted as evidence for a hot intergalactic medium with significant baryon content.   Subsequent X-ray imaging from large satellites (HEAO, ROSAT, Einstein Observatory) found a more complex situation, employing higher spatial resolution and 
spectra over a wide energy range (see \citet{Barcons1992} for a review).  These improved observations showed that the X-ray background comes from resolved non-thermal Galactic and extragalactic sources and diffuse soft X-ray emission from local hot interstellar gas. At high Galactic latitudes, a significant fraction of the diffuse background at 1 keV can be 
resolved into discrete sources.  However, in the 0.1 -- 0.3 keV band, the photon mean free path is limited to less than 100~pc at mean hydrogen density $n_{\rm H} \approx 1~{\rm cm}^{-3}$. Within the Local Bubble, which extends 60 -- 100 pc from the Sun \citep{Lallement2003} and is comprised of hot ($10^{6}$ K), low density ($n_{\rm H} \approx 0.01 cm^{-3}$) gas, the mean free path will be longer \citep{Snowden1990, Frisch2011}. While there is a 0.1 -- 0.3 keV contribution from charge exchange within the heliosphere \citep{Cravens2000, Lallement2004}, a significant fraction of soft X-rays is coming from the hot Local Bubble gas \citep{Galeazzi2014}.

This cosmic optical background (COB) has been the topic of several recent studies \citep{Driver2016, Mattila2017, SaldanaLopez2021} which have included deep images \citep{Zemcov_lorri_2017, Lauer2021, Lauer2022, Symons2023, Postman2024} taken with LORRI, the Long-Range Reconnaissance 
Imager \citep{Cheng2008, Weaver2020} onboard NASA's  New Horizons (NH) spacecraft.  
This background measures the red-shifted radiation produced by stars and gas in galaxies over the 
history of the Universe and serves as an important test of cosmological star-formation models.
\citet{Postman2024} used multiple LORRI images taken from far beyond zodiacal light interference in the distant Kuiper Belt, to measure the COB integrated 
from 0.4 -- 0.9 $\mu$m. That survey included 16 fields at high Galactic latitudes, selected to minimize scattered light from the Milky Way galaxy. These were augmented by 
eight calibration fields for diffuse Galactic light (DGL) and several auxiliary fields. 
The survey is free of the zodiacal light produced by sunlight scattered by interplanetary dust, 
and supersedes an earlier analysis \citep{Lauer2022} based on observations of one of the present 
fields. Isolating the COB contribution to the raw total sky levels measured in the fields required subtracting 
the remaining scattered light from bright stars and galaxies, the intensity from stars within the fields 
fainter than the photometric detection limit, and the DGL foreground.  The LORRI survey yielded a 
highly significant detection ($6.7 \sigma$) of the COB at $11.08\pm1.65$~nW~m$^{-2}$~sr$^{-1}$ 
at the pivot wavelength of $0.608~\mu$m.  The estimated integrated intensity from background 
galaxies, $8.17\pm1.18$~nW~m$^{-2}$~sr$^{-1}$, accounts for the majority of this signal, and is the most precise measurement of the COB to date. 

The current paper describes NH observations of the cosmic ultraviolet background (CUVB) radiation taken in parallel with the LORRI observations, but with small pointing offsets. The data were taken in the far-ultraviolet (FUV) by the Alice spectrograph \citep{Stern2008a} onboard the New Horizons spacecraft at a distance of 57~AU from the Sun, outside most of the interplanetary gas and dust that complicates such observations. Radiation in the FUV band (conventionally from 912~\AA\ to 2000~\AA) is important for studying a variety of astrophysical processes: massive star formation and ionized gas in the Galactic ISM;  heating of diffuse interstellar gas clouds by photoelectric emission from dust; controlling the interstellar atomic-to-molecular transition through photodissociation in the \Htwo\ Lyman and Werner bands; and the star-formation rate in galaxies over billions of years.  

Measurements of the CUVB surface brightness have a long history with a variety of instruments \citep[see reviews by][]{Bowyer1991, Henry1991, Murthyreview2009}.  These studies began with observations from sounding rockets \citep{Hayakawa1969, Henry1977, Anderson1979, Tennyson1988} and continued with  space-borne experiments aboard OAO-2 \citep{Lillie1976}, Apollo 17 \citep{Henry_ngp1978}, Voyager \citep{Murthy_voy, Murthyvoy_all}, and \GALEX\ \citep{Murthy_galex_data2010, Hamden2013, Akshaya2018, Akshaya2019, Chiang2019}. The CUVB has been found to be correlated with the amount of dust in the line of sight, usually represented by the color excess \ebv. Although this relationship saturates at low Galactic latitudes where the optical depth of the interstellar dust is high, there is a linear correlation of the CUVB and the \ebv\ at high Galactic latitudes where the optical depth of the dust is less than 1 \citep{Murthy_dustmodel2016}. This part of the CUVB is due to the scattering of starlight from hot stars in the Galactic Plane by the interstellar dust in the line of sight \citep{Jura1979}. The offsets, after subtraction of the dust-scattered light, represent any other isotropic emission, such as line emission from halo gas, the integrated light from unresolved galaxies, and other Galactic and extragalactic sources.

\begin{table*}[htbp]
\centering
\caption{Observations of the CUVB Offset at the Poles}
\label{tab:polar}
\begin{tabular}{lccl}
\hline
Reference & Wavelength (\AA) & Offset$^{a}$ & Instrument \\
\hline
\citet{Henry_ngp1978}& 1180 -- 1680 & 250 & Apollo 17\\
\citet{Anderson1979} & 1230 -- 1680 & $285 \pm 32$ & Rocket \\
\citet{Paresce1980} & 1350 -- 1550 & $<300$ & ASTP\\
\citet{Feldman_hotgas1981} & 1200 -- 1670 & 150 $\pm$ 50 & Rocket \\
\citet{Joubert1983} & 1690 & 300 -- 690 & D2B\\
\citet{Jakobsen1984} & 1590 & $<550$ & Rocket \\
& 1710 & $<900$ \\
\citet{Holberg1986} & 900 -- 1100 &  $< 200$ & Voyager\\
\cite{Onaka1991} & 1500 & 200 -- 300 & Rocket \\
\cite{Henry1993} & 1500 & 300 $\pm$ 100 & UVX\\
\cite{Witt1994} & 1500 & $300 \pm 80$ & DE-1 \\
\cite{Witt1997} & 1400 -- 1800& $160 \pm 50$ & FAUST \\
\cite{Schiminovich2001} & 1740& $200 \pm 100$ & NUVIEWS \\
\cite{Hamden2013} & 1565 & 300 & GALEX \\
\citet{Akshaya2018} & 1565& 240 -- 290 & GALEX\\
\citet{Akshaya2019} & 1565 & $240 \pm 18$ & GALEX\\
\hline
\multicolumn{4}{l}{$^{a}$ The offset is the remaining emission after subtraction of the dust-scattered light}\\
\multicolumn{4}{l}{in photon units (ph cm$^{-2}$ s$^{-1}$ sr$^{-1}$~\AA$^{-1}$).}
\end{tabular}
\end{table*}

There have been many observations of the CUVB near the Galactic Poles finding offsets of 230 -- 290 ph cm$^{-2}$ s$^{-1}$ sr$^{-1}$~\AA$^{-1}$ (Table \ref{tab:polar}). These units are commonly referred to as ``photon units" 
or ``continuum units". In the literature, integrated backgrounds are often converted to monochromatic fluxes,
$\nu I_{\nu} = \lambda I_{\lambda}$ in nW~m$^{-2}$~sr$^{-1}$, with the usual relation between  
flux distributions. Defining photon flux $\Phi_{\lambda} = I_{\lambda}/h\nu = ( \lambda I_{\lambda} / hc)$ 
at the central wavelength of the FUV band and taking care with \AA-to-cm conversion in the units, 
we find that 300 \photu\ = 5.96 nW~m$^{-2}$~sr$^{-1}$.   The integral of photon flux times energy
($hc/\lambda$) over the wavelength band can be expressed
\begin{equation}
    \int_{\lambda_1}^{\lambda_2}  \frac {hc}{ \lambda} \,  \Phi_{\lambda} \, d \lambda 
     \approx  hc \;  \bar {\Phi}_{\lambda}  \ln ( \lambda_2/ \lambda_1)  =
       \langle \lambda I_{\lambda} \rangle  \ln ( \lambda_2/ \lambda_1)   \; ,
\end{equation}  
where $ \bar {\Phi}_{\lambda}$ and $\langle \lambda I_{\lambda} \rangle $ are evaluated at 
band center.  By convention, most background surveys quote $\lambda I_{\lambda}$ without 
the factor  $\ln ( \lambda_2/ \lambda_1)$, an approximation that overestimates the actual width 
of the FUV band of \GALEX\ (440~\AA\ at  $\lambda_{\rm eff} \approx 1530$~\AA).    

The observed values of the offsets (Table \ref{tab:polar}) are considerably greater than the $73 \pm 16$ \photu\ expected from galaxy counts \citep{Driver2016}, with other suggested examples of 
possible FUV sources including the two-photon continuum produced from \HI\ $(2s \rightarrow 1s)$ 
emission produced in the warm ionized interstellar medium and low-velocity shocks \citep{Reynolds1992, Kulkarni2022, Kulkarni2023}, and the radiative decay of massive neutrinos \citep{Sciama1990} or other dark matter \citep{Kollmeier2014, Henry2015}. \citet{Kulkarni2022} suggested that much (41 -- 60 \photu) of the excess emission of 80 -- 230 \photu\ was due to two-photon emission arising in the Earth's atmosphere or in the interplanetary medium, as the earlier observations were generally made from low Earth orbit. We note that instrumental scattered light from off-axis stars may contribute to the observed signal but this will be minimized at the Galactic poles where there are few bright UV stars.

The spectral region between 912 -- 1200~\AA\ is much more difficult to observe because of internal scattering from the \lya\ line, either from the Earth's atmosphere or from the interplanetary medium. There has been only one observation at the Galactic poles using the \voyager\ ultraviolet spectrometers (UVS), which was only able to set an upper limit of about 200 \photu\ on the observed CUVB from 912 -- 1150~\AA\ \citep{Holberg1986}. Although the two \voyager\ UVS were able to detect the diffuse UV background in many parts of the sky, these observations were at the limit of the instrument's capabilities and could not significantly constrain the extragalactic background light (EBL) below 1200~\AA, especially because of the uncertain contribution due to internal scattering of interplanetary \lya.

The NH CUVB program was designed to measure the diffuse UV radiation field as viewed from the Kuiper Belt at a distance of 57 AU from the Sun, beyond the bulk of the emission from the Solar System. We will describe the analysis of the New Horizons Alice data and will discuss the measurement of the offsets. Those observations in the 1400 -- 1800~\AA\ band are consistent with earlier observations made from Earth orbit. Our determination of the offsets between 912 -- 1150~\AA\ are the first such in this band and are possible only because of the much lower contributions of the interplanetary hydrogen lines at this distance from the Sun.

\clearpage
\section{The CUVB Survey}

\subsection{The New Horizons Alice UV Spectrograph}

\begin{figure}
    \includegraphics[width=3in]{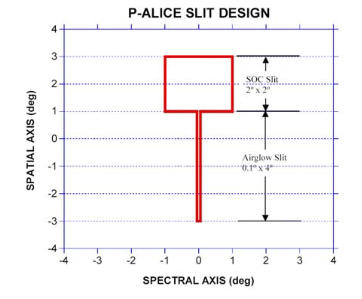}
    \caption{The Alice entrance aperture is a square Box on top of a narrow, rectangular Stem \citep{Stern2008}.}
    \label{fig:alice_slit}
\end{figure}

 The New  Horizons Alice spectrograph \citep{Stern2008} is a Rowland Circle spectrograph with spectral coverage from 520 – 1870~\AA, enabling direct measurement of the FUV  cosmic background over these wavelengths. The main airglow channel (AGC) has an aperture comprising two connected areas on the sky: a narrow ``Stem" with a field of view (FOV) of $0\adeg1\times  4\adeg0$ and a square ``Box" with a FOV of $2\adeg0 \times 2\adeg0$ (Fig. \ref{fig:alice_slit}). The full detector image is 1024 pixels in the spectral direction and 32 pixels in the spatial direction. The full slit illuminates roughly rows 6 -- 25 (zero-indexed) on the detector data array, with the Stem portion of the slit illuminating rows 6 -- 18 and the Box portion illuminating rows 19 -- 25, with row 18 serving as a transition between the two slit widths. Row 16 defines the instrument boresight. The full-width at half-maximum (FWHM) spectral resolution for point sources for the Stem and the Box is 1 -- 3~\AA\ and 5~\AA, respectively, and spatially each row is 0.3 deg along the slit. The Alice detector is an intensified Z-stack micro-channel plate (MCP) with a split coating of KBr (520 -- 1180 \AA) and CsI (1250 -- 1870 \AA) to cover the entire spectral range. The MCP was masked around the \lya\ line (1216~\AA) during the coating process to reduce sensitivity to that intense interplanetary line. The \lya\ line is approximately in the center of the spectral range with a FWHM (for aperture-filling diffuse sources) of $9 \pm 1.4$~\AA\ for the Stem and 172~\AA\ for the Box \citep{Stern2008a}. For the present survey, no CUVB measurements include detector regions directly illuminated by \lya; however, as we will show, much of the observed signal is from dark counts and instrumental scattering of the intense \lya\ line, which have to be modeled and subtracted from the data.

\begin{figure*}[bhtp]
\centering
\includegraphics[keepaspectratio,width=7.0 in]{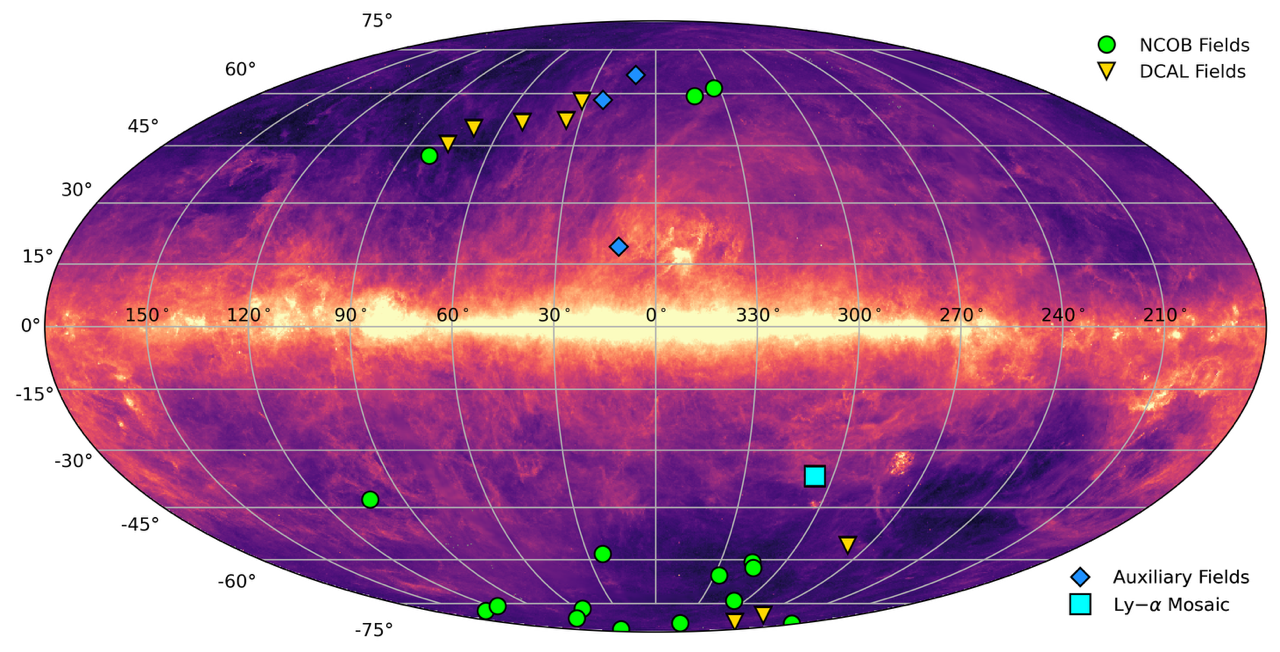}
\caption{The locations of the CUVB fields are shown on the IRIS full-sky $100 {~\rm\mu m}$ map \citep{MivilleDeschenes2005} in Galactic coordinates.  The auxiliary fields include the two shock fields and one molecular hydrogen fields.}
\label{fig:full_sky}
\end{figure*}

The AGC has a controllable aperture door which, when closed, permits an accurate measurement of the detector dark current.  We will detail the strategy for accurate measurement of the dark signal, which represents a significant fraction of the raw total signal of the on-sky observations, in $\S\ref{sec:obs}$. In this context, we also note that Alice has a special solar occultation channel (SOCC), which enabled observation of UV transmission spectra of Pluto's atmosphere. The SOCC is mounted at a $90^\circ$ angle to the AGC boresight, and has an optical path to the detector that is not shuttered. As it is intended for observation of the Sun, its throughput is a factor of 6400 less than the AGC, but it does represent a light leak into the instrument from the sky. As we detail in the next section, the spacecraft was oriented during the CUVB observations to place the SOCC port within the spacecraft shadow (see Fig. \ref{fig:full_sky}).

\subsection{The Survey Design and Field Selection}\label{sec:design}

The CUVB survey was designed in parallel with the NH COB survey, described by \citet{Postman2024}. In brief, they observed several fields with the New Horizons LORRI instrument to measure the COB intensity and calibrate the observations. Of these, 16 fields were selected for primary observation of the COB intensity; these were designated with the `NCOB' prefix. Eight additional fields, designated with the `DCAL' prefix, were selected to develop a DGL estimator when used in league with the NCOB fields.  Each of the LORRI NCOB and DCAL fields has a corresponding Alice field in the present survey to measure the CUVB intensity, albeit with small positional differences, as we detail below.  The CUVB program also includes four additional fields unique to the CUVB program for calibration purposes.

The overall geometry of the COB and CUVB surveys was specified by the trajectory of NH out of the solar system, the requirement that the Alice and LORRI apertures be positioned in the spacecraft shadow to avoid scattered sunlight affecting the background measurements, and avoidance of the dense regions of the Milky Way plane. The spacecraft trajectory in turn was specified by NH's primary mission of obtaining the first exploration of Pluto \citep{Stern2015} and then the Kuiper belt object Arrokoth \citep{Stern2019}. 

During the NH mission, Pluto, as seen from the Earth, was projected against the bulge of the Milky Way. This means that the ``anti-solar" hemisphere suitable for background measurements is roughly centered on the heart of our Galaxy. In detail, we selected fields with solar elongation angle (SEA) $>95^\circ$. While SEA~$>90^\circ$ would be sufficient to keep direct sunlight out of the instrument apertures, the spacecraft bulkhead in which the apertures are positioned supports other instruments that could potentially scatter sunlight into the apertures; SEA~$>95^\circ$ ensures that these are also shaded by the spacecraft \citep{Lauer2021}. Fields were selected with Galactic latitude $|b|>40^\circ$ to avoid dense stellar foregrounds and to minimize the contribution from dust-scattered starlight. Lastly, ecliptic latitudes were restricted to $|\beta|>15^\circ.$ While NH is not directly affected by zodiacal light, the FIR intensities that were used to select fields for low DGL are provided by maps made in Earth-space, and thus may incur larger errors near the ecliptic \citep{ Matsuura2011, Korngut2022, Carleton2022}.

The combination of these constraints left 4239 $\rm deg^2$ of sky available. \citet{Postman2024} randomly selected 60,000 positions within this area, and estimated the DGL contribution for each one using the IRIS 100$~\mu$m all-sky data \citep{MivilleDeschenes2005}.\footnote{\citet{Postman2024} developed an improved DGL estimator based on Planck 350$~\mu$m and 550$~\mu$m intensities, but used the \citet{zemcov_nh2018} DGL estimator at 100 \micron\ for the initial definition of the COB survey.} The 16 NCOB fields were selected to minimize estimated DGL and provide good angular coverage around the sky.  The fields were also selected to minimize scattered starlight in the LORRI field, although this was not a concern for the present Alice observations because there are so few bright stars in the UV, particularly at high Galactic latitudes.

The eight DCAL fields were selected to perform an improved self-calibration of the relation of FIR intensity to UV DGL in combination with the NCOB fields.  As such, they were located to cover fields with progressively higher $100~\mu$m surface brightness, up to a limit of $\sim 3~{\rm MJy~sr^{-1}}$. This limit was selected to avoid dust optical depths large enough that non-linear behavior between the FIR intensity and scattered light amplitude might occur.

The final coordinates of each Alice NCOB and DCAL field were adjusted by up to $\sim1\adeg5$ compared to the corresponding LORRI field to minimize the presence of UV-bright stars within the Alice FOV. The orientation of the fields were also rolled about the Alice optical axis to bring the SOCC aperture within the spacecraft shadow.

\begin{deluxetable*}{lrrccrrrrcc}
\tabletypesize{\scriptsize}
\tablecolumns{11}
\tablewidth{0pt}
\tablecaption{Survey Field Centers and Observations}
\tablehead{
\multicolumn{9}{c}{~~}\\
\multicolumn{1}{c}{} &
\multicolumn{2}{c}{R.A.~~(J2000)~~Dec.} &
\multicolumn{1}{c}{} &
\multicolumn{1}{c}{} &
\multicolumn{2}{c}{Stem} &
\multicolumn{2}{c}{Box} &
\multicolumn{1}{c}{Stem} &
\multicolumn{1}{c}{Box} \\
\multicolumn{1}{c}{Field ID} &
\multicolumn{1}{c}{(deg)} &
\multicolumn{1}{c}{(deg)} &
\multicolumn{1}{c}{UT Date} &
\multicolumn{1}{c}{MET (s)} &
\multicolumn{1}{c}{$l$ (deg)} &
\multicolumn{1}{c}{$b$ (deg)} &
\multicolumn{1}{c}{$l$ (deg)} &
\multicolumn{1}{c}{$b$ (deg)}&
\multicolumn{1}{c}{E(B-V)}&
\multicolumn{1}{c}{E(B-V)}
}
\startdata
NHTF01 & 359.559 &$-23.654$& 2021-09-24 & 0494772718 &  57.38 &$-76.40$&  51.93 &$-78.98$& 0.013 & 0.018 \\
NCOB01 &   6.415 &$-56.809$& 2023-09-13 & 0556922877 & 313.30 &$-61.03$& 318.88 &$-60.12$& 0.017 & 0.013\\
NCOB02 &   8.410 &$-54.967$& 2023-08-30 & 0555729417 & 313.49 &$-63.51$& 312.02 &$-60.70$& 0.016 & 0.016\\
NCOB03 &   0.443 &$-47.405$& 2023-08-21 & 0554939877 & 325.99 &$-65.55$& 331.63 &$-63.94$& 0.014 & 0.014\\
NCOB04 &  12.983 &$-41.939$& 2023-08-20 & 0554877517 & 305.36 &$-74.57$& 311.98 &$-72.34$& 0.012 & 0.011\\
NCOB05 &   8.762 &$-23.312$& 2023-08-28 & 0555543697 &  84.36 &$-86.79$& 116.06 &$-87.85$& 0.012 & 0.013\\
NCOB06 &  11.659 &$-35.055$& 2023-09-12 & 0556828417 & 336.32 &$-84.42$& 324.18 &$-81.82$& 0.015 & 0.011\\
NCOB07 &  20.604 &$-24.683$& 2023-08-27 & 0555482077 & 184.71 &$-83.37$& 209.75 &$-84.38$& 0.018 & 0.015\\
NCOB08 & 334.996 &$-25.720$& 2023-08-30 & 0555667297 &  24.99 &$-58.13$&  19.55 &$-58.38$& 0.014 & 0.011\\
NCOB09 &   6.231 &$-20.811$& 2023-08-29 & 0555608978 &  81.04 &$-80.57$&  66.93 &$-82.49$& 0.014 & 0.013\\
NCOB10 &  18.288 &$-16.543$& 2023-08-27 & 0555420517 & 135.27 &$-77.06$& 137.93 &$-79.83$& 0.015 & 0.014\\
NCOB11 &   9.843 &$-13.184$& 2023-08-26 & 0555358257 & 115.88 &$-75.09$& 114.03 &$-77.89$& 0.017 & 0.013\\
NCOB12 & 204.187 &$  2.155$& 2023-08-17 & 0554554437 & 331.73 &$ 60.86$& 334.62 &$ 63.38$& 0.025 & 0.025\\
NCOB13 & 211.930 &$  2.394$& 2023-08-17 & 0554547508 & 341.55 &$ 58.37$& 344.92 &$ 60.66$& 0.030 & 0.024\\
NCOB14 & 356.932 &$ 17.336$& 2023-08-20 & 0554837698 & 103.05 &$-42.59$&  99.44 &$-43.73$& 0.073 & 0.058\\
NCOB15 & 247.924 &$ 52.331$& 2023-08-13 & 0554209418 &  79.25 &$ 42.35$&  83.04 &$ 41.76$& 0.025 & 0.021\\
DCAL01 &  20.257 &$-30.111$& 2023-08-20 & 0554845297 & 260.77 &$-83.94$& 276.07 &$-81.65$& 0.022 & 0.018\\
DCAL02 &  24.026 &$-34.454$& 2023-09-12 & 0556796457 & 260.38 &$-80.80$& 264.89 &$-77.99$& 0.023 & 0.017\\
DCAL03 & 236.278 &$ 43.411$& 2023-08-13 & 0554247327 &  69.05 &$ 49.90$&  73.40 &$ 49.32$& 0.020 & 0.017\\
DCAL04 & 240.679 &$ 47.617$& 2023-08-13 & 0554216007 &  74.83 &$ 45.48$&  78.83 &$ 44.88$& 0.014 & 0.026\\
DCAL05 & 239.889 &$ 35.271$& 2023-08-14 & 0554300848 &  51.12 &$ 51.52$&  55.74 &$ 51.58$& 0.033 & 0.027\\
DCAL06 &  37.452 &$-56.481$& 2023-09-13 & 0556916078 & 277.65 &$-56.38$& 281.09 &$-54.25$& 0.041 & 0.039\\
DCAL07 & 228.234 &$ 22.238$& 2023-08-15 & 0554367758 &  30.54 &$ 57.45$&  35.77 &$ 58.13$& 0.048 & 0.053\\
DCAL08 & 234.766 &$ 22.599$& 2023-08-14 & 0554336068 &  34.12 &$ 51.79$&  38.66 &$ 52.46$& 0.061 & 0.055\\
LYACAL &  10.095 &$-35.029$& 2023-08-22 & 0555002780 & 306.70 &$-35.62$& 306.35 &$-38.47$& 0.160 & 0.076 \\
H2\_NE  & 257.709 &$ -9.334$& 2023-09-06 & 0556334397 &  10.34 &$ 19.53$&  13.01 &$ 18.14$& 0.747 & 0.594\\
SHOCK1 & 226.189 &$ 17.788$& 2023-08-15 & 0554428408 &  21.55 &$ 57.64$&  26.63 &$ 58.65$& 0.032 & 0.042\\
SHOCK2 & 215.952 &$ 15.941$& 2023-08-16 & 0554488058 &   8.33 &$ 65.37$&  14.50 &$ 66.75$& 0.020 & 0.020\\
\enddata
\tablecomments{The R.A.\ and Dec values refer to the coordinates of the overall Alice boresight, while Galactic coordinates are given for the centers of the separate Stem and Box apertures. All coordinates are in degrees.  MET is the mission elapsed time in seconds (from the launch on 14:00 ET on Jan 19, 2006) of the first image in each field sequence. NCOB fields are the primary fields for measuring the COB intensity. NHTF01 is the test of the NCOB field selection and observational strategy published initially in \citet{Lauer2022} and reanalyzed in \citet{Postman2024}. DCAL fields are for DGL calibration. E(B - V) values are the mean values in the Stem and the Box, respectively, from \citet{PlanckDust2016} with a typical variation of 5 millimagnitudes.}
\end{deluxetable*}\label{tab:coords}

Lastly, four fields unique to the CUVB program were defined to improve calibration of the Alice instrument, or to augment analysis of the observations.  Two fields, designated SHOCK1 and SHOCK2, were defined to observe UV emission from highly shocked gas associated with the Fermi/eROSITA bubbles at Galactic latitudes $57\adeg6$ and $65\adeg4,$ respectively.
H2\_NE is a low-Galactic latitude field selected to observe possible molecular hydrogen (H$_{2}$) fluorescent emission. LYACAL is a re-observation of a mosaic of a contiguous area of Box fields first observed with Alice in 2007 when the spacecraft was $\sim8$ AU from the Sun, with the goal of characterizing the Ly$\alpha$ scattering function of the Alice spectrograph (see Section \ref{Sec:template}). Differencing the 2023 and 2007 datasets removes invariant astrophysical sources common to both, isolating the Solar Ly$\alpha$ emission, which has decreased as the spacecraft traveled to 57 AU \citep{Murthy_voy, Gladstone_lya2018}.

The coordinates and observation dates for the CUVB fields are listed in Table~\ref{tab:coords}. The MET (mission elapsed time) identifiers of the first observation of each field are also given. Figure~\ref{fig:full_sky} shows the field distribution on the sky with respect to the IRIS $100~\mu$m map \citep{MivilleDeschenes2005}. 

\subsection{The CUVB Observations}\label{sec:obs}

\begin{figure*}
    \includegraphics[width=7in]{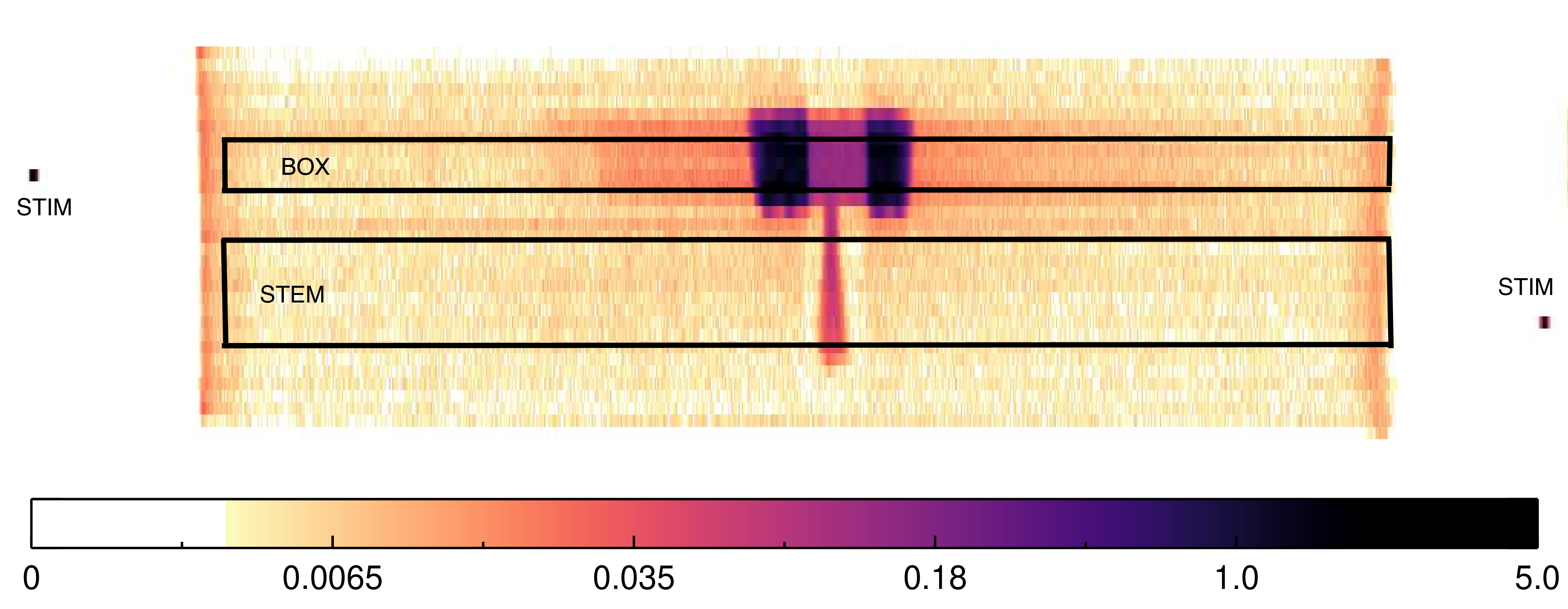}
     \caption{Detector image of one of our diffuse background observations, with the instrumental STIM pulses seen on either side of the image. The scale is in units of counts pixel$^{-1}$ s$^{-1}$. Wavelength increases to the right with the illuminated region spanning the range from 520 -- 1870~\AA.  \lya\ is at the center of the image with much of the background due to instrumental scattering and dark counts. The Stem is identified in the lower part of the image, and the Box in the upper part.}
    \label{fig:alicedata}
\end{figure*}

A sequence of individual exposures, using the Alice Histogram Imaging Mode (HIM), was taken at each field. The histogram data consists of FITS \citep{FITS2010} images with the counts integrated over the entire observation (Fig. \ref{fig:alicedata}). The detector image is 1024 pixels in the spectral direction and 32 pixels in the spatial direction, with the extent of the Stem and the Box shown in Fig. \ref{fig:alice_slit}. The Stem
spectrum is extracted from rows 6 to 15 (inclusive) and the Box spectrum is extracted from rows 20 to 24 to minimize any vignetting effects near the edge and Stem/Box transition of the slit. The Ly$\alpha$ line is approximately in the center of the spectral range with a full-width at half-maximum (FWHM) of 9~\AA\ for the Stem and 172~\AA\ for the Box. 

An important consideration for understanding the region of the sky observed in any observation is that the pointing of NH is controlled by mono-propellant thrusters, rather than the precise reaction wheels used in spacecraft such as HST or JWST.  For Alice observations, the integrations are conducted with a $0\adeg14$ guidance deadband, which means that the Alice aperture will wander around the sky within this distance from the commanded pointing. As this is larger than the width of the Alice stem, this means that the area of the sky at each pointing that it samples is at least twice its width.  For precise use, spacecraft telemetry is available showing the trajectory of the Alice aperture over any observation. 

A major concern was the accurate and precise measurement of the Alice dark current, which contributes a significant fraction of the total signal in the sky exposures.  To this end, Alice observations commenced only after a five-hour thermal stabilization period following the activation of the instrument.  This interval was considered adequate to counter the small temperature-dependent variations in Alice sensitivity seen on shorter timescales following power-on.

\begin{table*}[t]
\caption{Dark Observations with Alice
\label{tab:darklog}}
\begin{tabular}{lclllll}
\hline\hline
Year & NExp$^a$ &  Exp.$^{b}$ & D$_{STEM}^{c}$ & E$_{STEM}^{d}$ & D$_{BOX}^{e}$ & E$_{BOX}^{f}$\\
\hline
 2007 &   3  & 10,800&  0.00368 &  0.00018 &  0.00442 &  0.00029\\
 2008 &   3  & 10,800 &  0.00365 &  0.00018 &  0.00434 &  0.00028\\
 2010 &   3  & 10,720&  0.00383 &  0.00019 &  0.00436 &  0.00028\\
 2012 &   3  & 10,720 &  0.00372 &  0.00019 &  0.00417 &  0.00028\\
 2014 &   3  & 10,720&  0.00352 &  0.00018 &  0.00393 &  0.00027\\
 2021 &	 9	& 32,400 &  0.00395 &  0.00011 &  0.00472 &  0.00017\\
 2023 & 192  & 691,200&  0.00369 &  0.00002 &  0.00417 &  0.00003\\
\hline
\multicolumn{7}{l}{$^a$ Number of independent exposures.}\\
\multicolumn{7}{l}{$^b$ Cumulative exposure time in seconds.}\\
\multicolumn{7}{l}{$^c$ Mean dark in Stem.}\\
\multicolumn{7}{l}{$^d$ Mean error in Stem.}\\
\multicolumn{7}{l}{$^e$ Mean dark in Box}\\
\multicolumn{7}{l}{$^f$ Mean error in Box.}\\
\multicolumn{7}{l}{$^{c -- f}$ From 912 - 1800 \AA\ excluding \lya.}\\
\multicolumn{7}{l}{$^{c -- f}$ Units of counts s$^{-1}$ pixel$^{-1}$.}

\end{tabular}
\end{table*}

The standard observing sequence for any field was to start with a 3600 s dark exposure (aperture door closed), followed by several sets of a 3600 s sky and 3600 s dark exposure pairs.  This way the individual sky exposures are always interlaced between dark exposures, allowing any slow drift in dark current to be tracked.  For the NCOB and auxiliary fields eight individual sky exposures of 3600 s each were obtained for a 28,800 s total exposure, again interlaced between nine dark exposures.  For the DCAL fields, four 3600s sky exposures, for a total of 14,400 s, were interlaced between five dark exposures. The dark observations taken during the Alice mission are listed in Table \ref{tab:darklog} and will be discussed in Section \ref{sec:dark}.

\section{Data Analysis}

\subsection{Overview}

We used an identical data analysis procedure for the entire data set of CUVB observations. We began with the Alice histogram data from the KEM2 mission \citep{AliceUV_2025}, archived at the NASA Planetary Data System (\url{https://pds-smallbodies.astro.umd.edu/data_sb/missions/nh-kem/index.shtml}). The primary contributors to the signal are instrumental dark current, largely due to fast particles from the onboard radioisotope thermoelectric generator (RTG), and internal scattering of the intense \lya\ line across the detector. We describe the identification and subtraction of these two components of the overall signal in the following sections. The calculations were done in count-space to avoid biasing by the calibration curve.

\subsection{Dark Counts \label{sec:dark}}

\begin{figure}
    \includegraphics[width=3.5in]{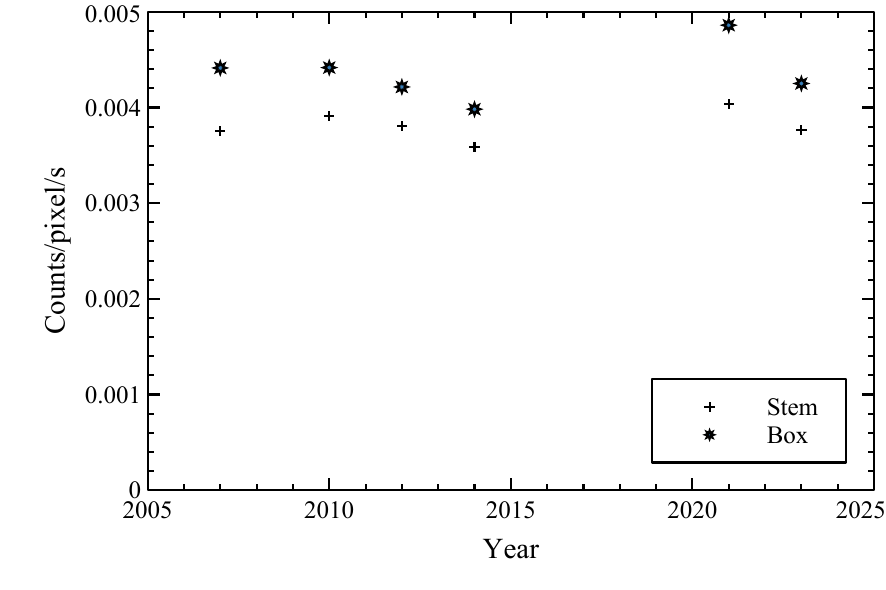}
    \caption{The average dark count rate over the Stem (+) and the Box (*) as a function of the year in which the observations were taken (Table \ref{tab:darklog}). The error bars were smaller than the data points.}
    \label{fig:darkyear}
\end{figure}

As discussed above, measurements of the dark count with the aperture door closed have been taken since the beginning of observations in 2007, with a systematic effort to interleave observations of the sky with dark measurements in the astrophysical program in 2023 (Table \ref{tab:darklog}). There is some variation in the dark rate as a function of observation year (Fig. \ref{fig:darkyear}), and we have used the dark count of the appropriate year in our dark subtraction. The errors in the dark spectrum were calculated assuming photon statistics (square root of the total number of counts) and have been taken into account in the error analysis.

\begin{figure}
    \includegraphics[width=3.5in]{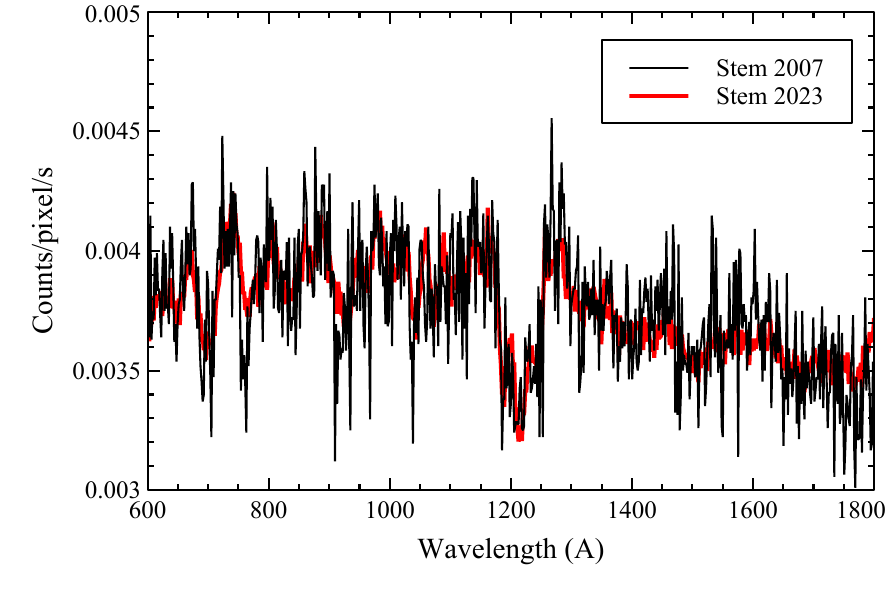}
    \caption{Spectrum of the dark counts in the Stem from 2007 (black line) and 2023 (red line). Note that there is less noise in 2023 because of the length of the dark observations (Table \ref{tab:darklog}).}
    \label{fig:darkstem}
\end{figure}

\begin{figure}
    \includegraphics[width=3.5in]{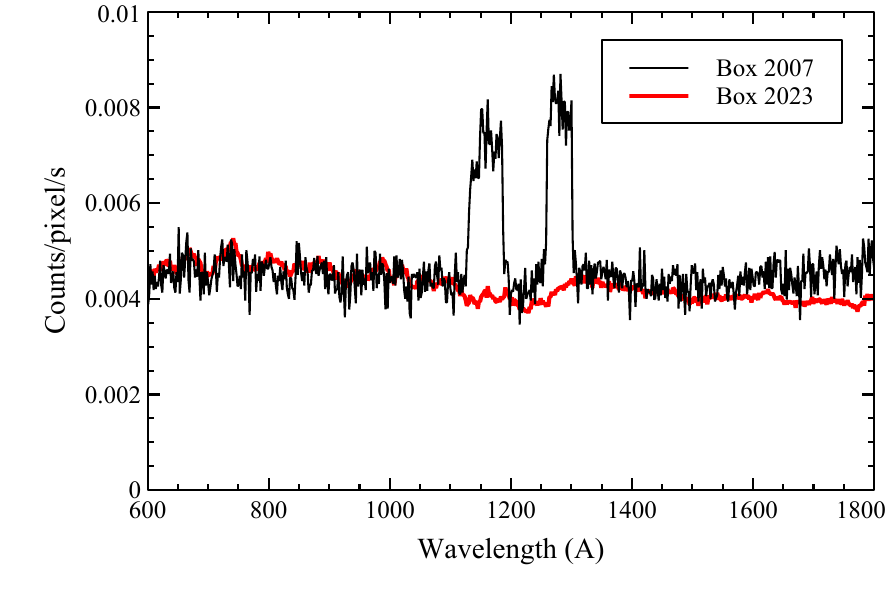}
    \caption{Spectrum of the dark counts in the Box from 2007 (black line) and 2023 (red line).}
    \label{fig:darkbox}
\end{figure}

The shape of the dark spectrum is constant in the Stem, with no difference in the individual spectra from 2007 to 2023 (Fig. \ref{fig:darkstem}). The Box data from early in the mission (2007) show significant \lya\ contamination (Fig. \ref{fig:darkbox}) that affects the region around 1216~\AA\ and at longer wavelengths, perhaps due to contamination from the solar occultation channel (SOCC). The SOCC had a solar elongation angle of about 17 degrees during the dark exposures in 2007 and did not have a door that could be closed. We took care to ensure that the SOCC was pointed to a dark patch of sky in the 2023 observations and, indeed, the dark Box spectra from 2023 show no evidence of contamination at \lya\ (Fig. \ref{fig:darkbox}).

\subsection{\lya\ Scattering Matrix} \label{Sec:template}

\begin{table*}[t]
\caption{\lya\ Template
\label{tab:templatelog}}
\begin{tabular}{lllllll}
\hline\hline
Year & l$^a$ & b$^b$ & N$^c$ &  Exp. Time (s) & STEM$^d$ & BOX$^e$\\
\hline
 2007  & 306.0 & -36.8 &    6  &    32,400 & 1170 & 953\\
 2023  & 306.0 & -36.7 &   10  &    36,000 & 1130 & 879\\
\hline
\multicolumn{7}{l}{$^a$ Mean Galactic longitude of boresight.}\\
\multicolumn{7}{l}{$^b$ Mean Galactic latitude of boresight.}\\
\multicolumn{7}{l}{$^c$ Number of exposures.}\\
\multicolumn{7}{l}{$^d$ Mean \galex\ surface brightness (\photu) in Stem.}\\
\multicolumn{7}{l}{$^e$ Mean \galex\ surface brightness (\photu) in Box.}\\
\end{tabular}
\end{table*}

\begin{figure*}
    \includegraphics[width=6in]{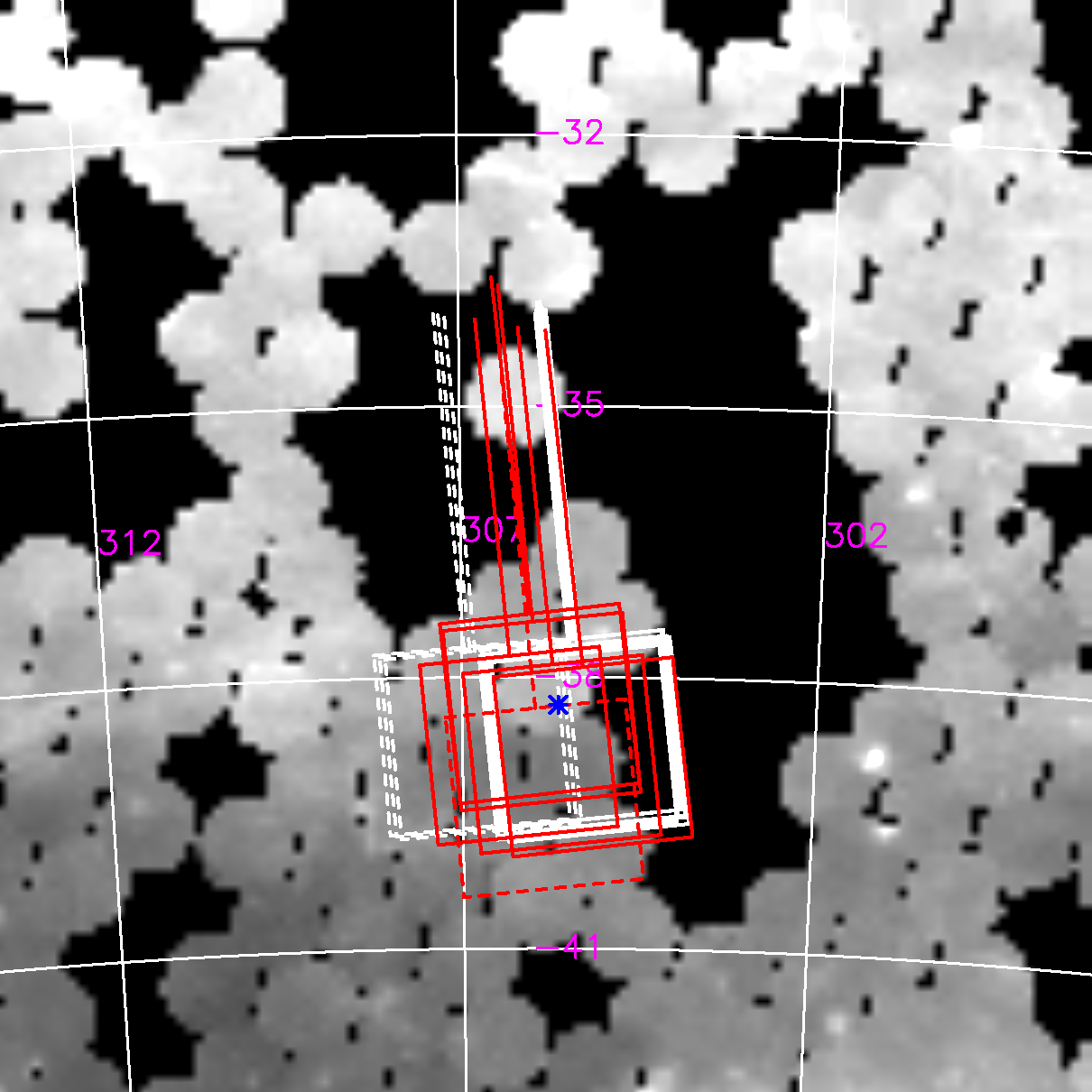}
    \caption{Diffuse FUV image from \galex\ \citep{Murthy2014apj} in Galactic coordinates with the Alice Box and Stem observations from 2007 (red) and 2023 (white) superimposed. The dashed observations did not include the star HIP 648  (blue star) in the Box and were not used in the derivation of the Box template. Black areas show where there were no \galex\ diffuse data.}
    \label{fig:lya_map}
\end{figure*}

In most cases, the strongest source of diffuse emission in the sky is the resonant scattering of \lya\ photons from the Sun by interplanetary hydrogen atoms. The Alice detector was masked around 1216~\AA\ during the coating process to reduce the number of counts due to \lya\ \citep{Stern2008a} but, despite this, internal scattering of the \lya\ photons contaminates much of the spectrum. Although the \lya\ scattering function was characterized during the ground calibration, it was difficult to simulate an aperture-filling diffuse field in the lab and the data were not well fitted by the ground scattering function.

The strength of the \lya\ line and, hence, the scattered light drops rapidly as a function of distance from the Sun \citep{Murthy_voy, Gladstone_lya2018}. Alice re-observed a long observation of the blank sky from 2007 (Table \ref{tab:templatelog}) to isolate the scattered \lya\ in the spectrum. The strength of the \lya\ line dropped by a factor of 3.5 between 2007, when the distance of NH from the Sun was about 8 AU, and 2023, when the distance was about 57 AU. There were 6 observations taken of the sky in 2007 and 10 in 2023, distributed as shown in Fig. \ref{fig:lya_map}. There was little variation in the diffuse background in the Stem observations, but there was a 7.2 magnitude A3 V star (HIP 648) in some, but not all, of the Box observations. The geometry of the observations was such that we obtained the maximum signal-to-noise when we used those observations (5 observations in 2007 and 7 in 2023) that included the star in deriving the \lya\ scattering function for the Box.

\begin{figure}
    \includegraphics[width=3.5in]{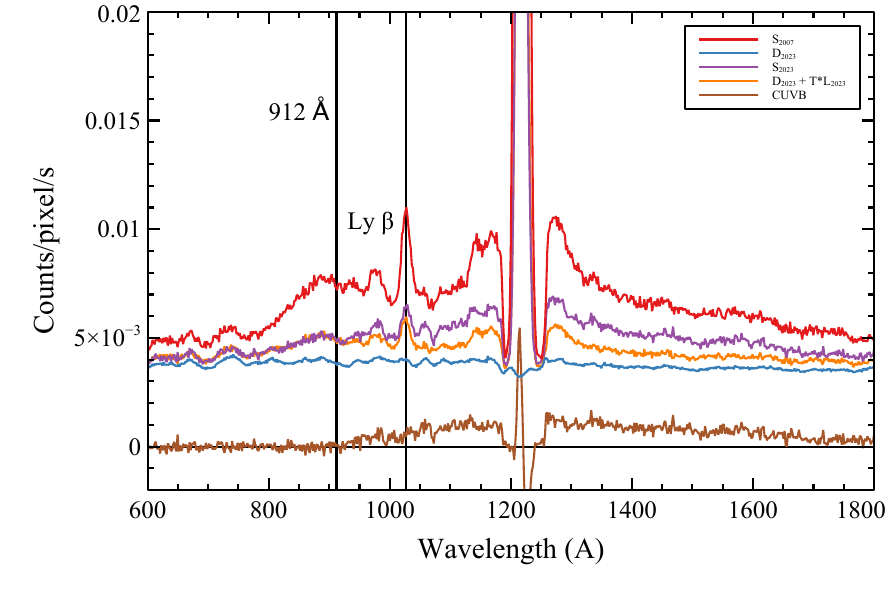}
    \caption{Different components of the template creation for the Stem. Symbols in the key are as defined in Eq. \ref{eq_stem_scatter}. The \lya\ scattering matrix is given by the difference between the observed spectrum in 2007 ($S_{2007}$) and 2023 ($S_{2023}$) and has to be scaled to the observed counts in the \lya\ line. The CUVB is the resultant after subtraction of the scaled scattering matrix and the dark current ($D_{2023} + T \times L_{2007}$)}
    \label{fig:lya_template}
\end{figure}

\begin{figure}
    \includegraphics[width=3.5in]{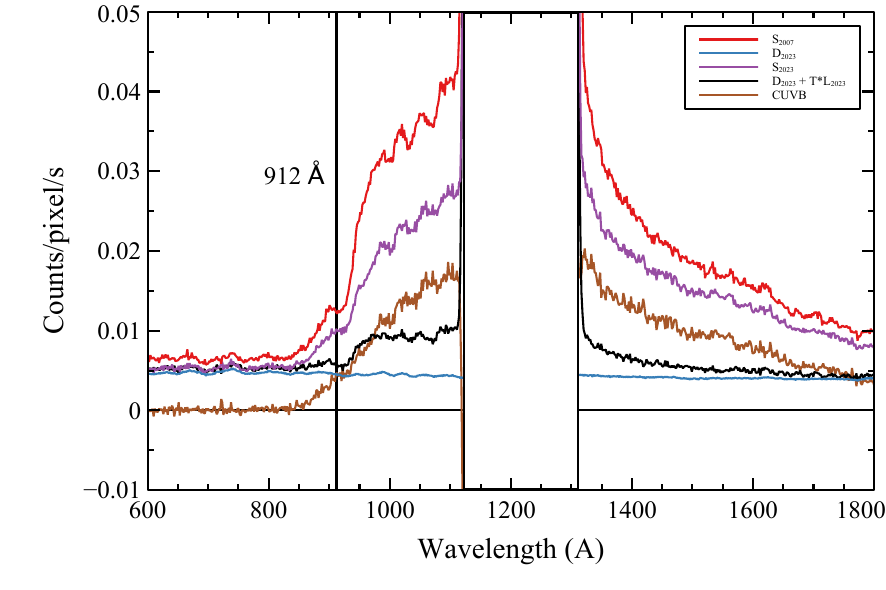}
    \caption{Different components of the template creation for the Box. Symbols in the key are as defined in Eq. \ref{eq_stem_scatter}. See caption for Fig. \ref{fig:lya_template}.}
    \label{fig:lya_template_box}
\end{figure}

The CUVB is the same in both sets of observations and the scattering template is simply:
\begin{equation}
    T = \frac{(S_{2007} - D_{2007}) - (S_{2023}- D_{2023})}{(L_{2007} - L_{2023})}
    \label{eq_stem_scatter}
\end{equation}
where 
\begin{itemize}
    \item $T$ is the \lya\ scattering template (to be derived from the data).
    \item $S_{2007}$ and $S_{2023}$ are the total signal in 2007 and 2023, respectively.
    \item $L_{2007}$ and $L_{2023}$ are the respective counts under the \lya\ line, calculated after subtraction of the dark counts.
    \item $D_{2007}$ and $D_{2023}$ are the dark spectrum in 2007 and 2023, respectively. The dark spectrum in the Stem was the same in both the 2007 observations and the 2023 observations and we used $D_{2023}$ for both sets of observations. 
\end{itemize}
The difference between the scaled \lya\ template ($T$) and the actual spectrum ($S$) is the CUVB (DGL + EBL):
\begin{equation}
    CUVB = (S - D) - L*T.
    \label{CUVB_eq}
\end{equation}
We have shown the different components of the observed spectrum in Fig. \ref{fig:lya_template} for the Stem and Fig. \ref{fig:lya_template_box} for the Box.

\subsubsection{CUVB \& Error Analysis \label{sec:CUVB}}

\begin{figure}
    \includegraphics[width=3.5in]{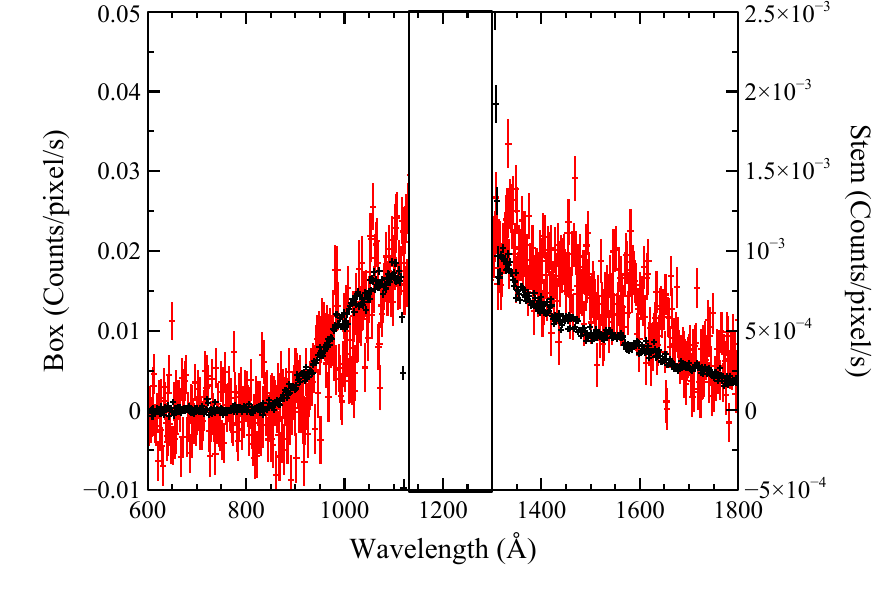}
    \caption{Derived CUVB in the Stem (red) and Box (black), plotted as $1\sigma$ error bars, where the derivation of the errors is discussed in Section \ref{sec:CUVB}. The solid angle subtended by the Stem is $\frac{1}{20}$ that of the Box and this is reflected in the observed count rate. We have not shown the region immediately around the \lya\ line where the subtraction of the intense \lya\ line is uncertain.}
    \label{fig:CUVB}
\end{figure}

\begin{table}[t]
\caption{Observed Surface Brightness in Box
\label{tab:nhbkgds}}
\begin{tabular}{llll}
\hline\hline
 Target & E(B - V)$^{a}$& S$_{1000}^{b}$ & S$_{1500}^{c}$ \\
\hline
NCOB04 & 0.011 & $ 245 \pm   15$ & $ 307 \pm   37$ \\
NCOB06 & 0.011 & $ 204 \pm   15$ & $ 279 \pm   36$ \\
NCOB08 & 0.011 & $ 204 \pm   14$ & $ 315 \pm   36$ \\
NCOB05 & 0.013 & $ 222 \pm   15$ & $ 318 \pm   37$ \\
NCOB11 & 0.013 & $ 221 \pm   15$ & $ 266 \pm   37$ \\
NCOB09 & 0.013 & $ 241 \pm   15$ & $ 361 \pm   37$ \\
NCOB01 & 0.013 & $ 268 \pm   15$ & $ 345 \pm   37$ \\
NCOB03 & 0.014 & $ 335 \pm   16$ & $ 398 \pm   38$ \\
NCOB10 & 0.014 & $ 224 \pm   15$ & $ 295 \pm   37$ \\
NCOB07 & 0.015 & $ 270 \pm   15$ & $ 441 \pm   39$ \\
NCOB02 & 0.016 & $ 263 \pm   15$ & $ 358 \pm   37$ \\
DCAL03 & 0.017 & $ 287 \pm   20$ & $ 496 \pm   51$ \\
DCAL02 & 0.017 & $ 272 \pm   20$ & $ 376 \pm   48$ \\
NHTF01 & 0.018 & $ 276 \pm   16$ & $ 993 \pm   47$ \\
DCAL01 & 0.018 & $ 311 \pm   20$ & $ 395 \pm   49$ \\
SHOCK2 & 0.020 & $ 314 \pm   16$ & $ 376 \pm   39$ \\
NCOB15 & 0.021 & $ 252 \pm   16$ & $ 437 \pm   40$ \\
NCOB13 & 0.024 & $ 357 \pm   16$ & $ 421 \pm   40$ \\
NCOB12 & 0.025 & $ 337 \pm   17$ & $ 365 \pm   40$ \\
DCAL04 & 0.026 & $ 291 \pm   20$ & $ 464 \pm   51$ \\
DCAL05 & 0.027 & $ 373 \pm   21$ & $ 566 \pm   52$ \\
DCAL06 & 0.039 & $ 387 \pm   21$ & $ 537 \pm   51$ \\
SHOCK1 & 0.042 & $ 342 \pm   16$ & $ 555 \pm   41$ \\
DCAL07 & 0.053 & $ 465 \pm   22$ & $ 776 \pm   56$ \\
DCAL08 & 0.055 & $ 408 \pm   21$ & $ 762 \pm   55$ \\
NCOB14 & 0.058 & $ 344 \pm   16$ & $ 620 \pm   40$ \\
LYACAL & 0.078 & $ 486 \pm   38$ & $1055 \pm  100$ \\
H2\_NE & 0.594 & $1084 \pm   20$ & $3642 \pm   65$ \\
\hline
\hline
\multicolumn{4}{l}{$^a$ Mean Planck E(B - V) in Box (mag).}\\
\multicolumn{4}{l}{$^b$ Surface brightness (912 -- 1150 \AA) in \photu.}\\
\multicolumn{4}{l}{$^c$ Surface brightness (1400 -- 1800 \AA) in \photu.}\\

\end{tabular}
\end{table}

The CUVB (DGL + EBL) may be identified with the residual after subtracting the \lya\ template and the dark count. We have plotted the CUVB for the Box (Note that the CUVB for the Box includes a small contribution from HIP 648 at the longest wavelengths) and the Stem as 1$\sigma$ error bars in Fig. \ref{fig:CUVB} where the error bars are calculated as follows:
\begin{enumerate}
    \item The error in the dark counts is the square root of the total number of counts in each pixel.
    \item The error in each of the two observations (2007 and 2023) is the square root of the number of counts in each pixel.
    \item The error in the \lya\ template is the error in each of the two observations added in quadrature.
    \item The final error in the CUVB is the square root of the sum of the squares of the errors in the observation, in the scaled template, and in the dark counts.
 \end{enumerate}
The error bars are much smaller for the Box than the Stem because the Box is much larger and is, therefore, more sensitive to diffuse radiation, at the cost of spatial and spectral resolution.
We have applied the same dark and template subtraction to all of our observations and will discuss the results below.

\section{Results}

\subsection{Overview}

\begin{figure*}
    \includegraphics[width=7in]{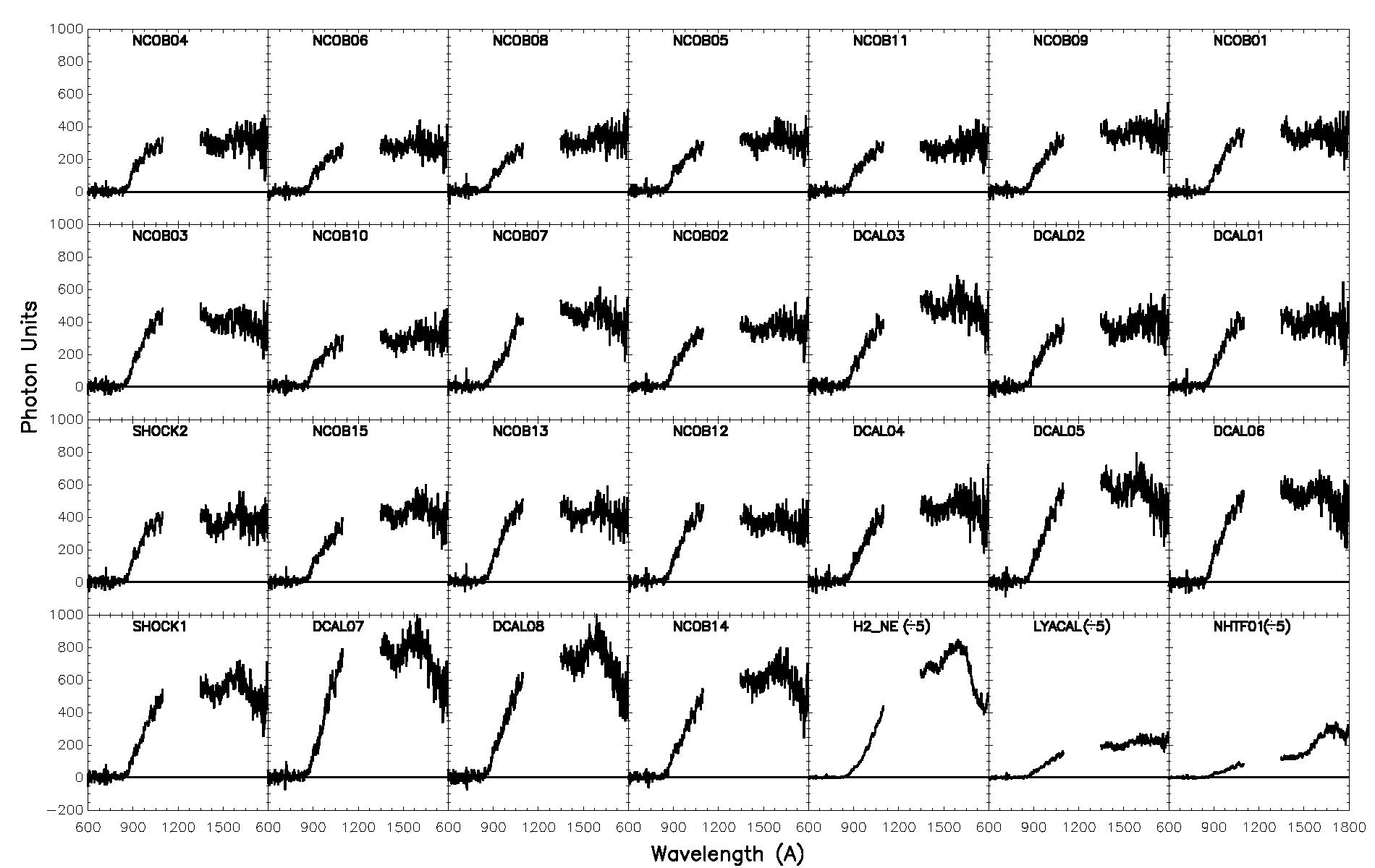}
    \caption{Box spectrum for each of the fields plotted as a function of wavelength (600 -- 1800~\AA). The spectra are ordered by the mean Planck E(B~-~V) in the Box, except for the last three spectra, and are labeled as per Table \ref{tab:coords}. The surface brightness in the last three spectra have been divided by 5 for plotting purposes and are not used in the analysis (see text). The data are available in electronic format.}
    \label{fig:box}
\end{figure*}

We have calculated the CUVB in each of the Stem and Box observations separately using Equation \ref{CUVB_eq}. The calculations were performed in counts-space (counts s$^{-1}$ pixel$^{-1}$) and then converted into \photu\ using a calibration determined through a comparison with \galex\ observations (described in the next section). The spectra are plotted for the Box (Fig. \ref{fig:box}) with the mean surface brightness in each band tabulated in Table \ref{tab:nhbkgds}.
We have not shown the spectra from the Stem because of their poor signal-to-noise ratio. Both the Stem and the Box data are available in electronic format.

We have excluded the following observations from our analysis in the rest of this work:
\begin{enumerate}
    \item ``H2\_NE'' was located at a low Galactic latitude to search for molecular hydrogen fluorescence and its modeling is complex with multiple components.
    \item There was a 4\textsuperscript{th} magnitude (V magnitude) B5 star near the Stem in ``DCAL04''.
    \item There was a 6\textsuperscript{th} magnitude (V magnitude) A1 star in the Box in ``NHTF01''.
    \item We exclude the template observation (``LYACAL'').
\end{enumerate}

\subsection{Comparison with \galex}
\begin{figure}
    \includegraphics[width=3.5in]{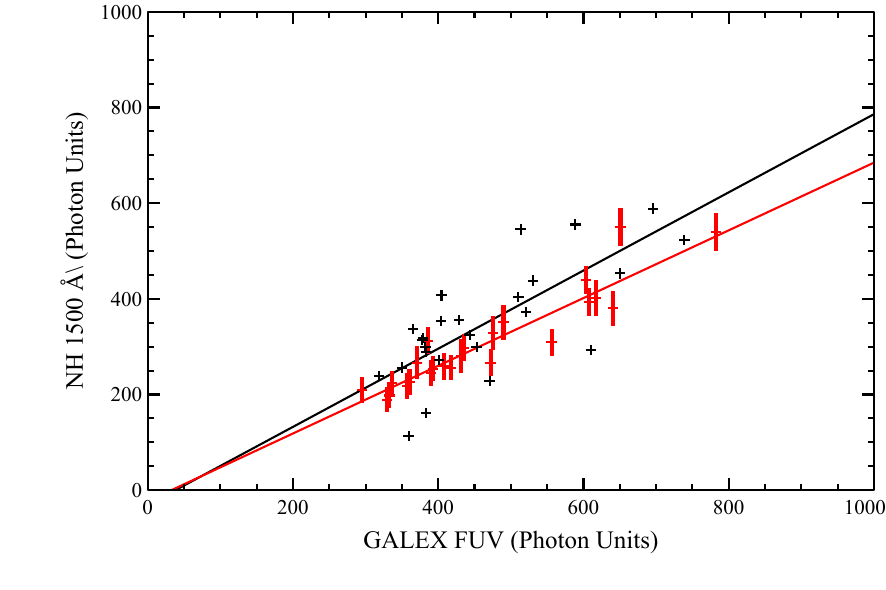}
    \caption{Stem (black) and Box (red) Alice surface brightness between 1350~\AA\ and 1800~\AA\ compared to the mean \galex\ surface brightness in the FUV band. The best-fit lines are $NH = 0.82G - 31.2$ for the Stem (black line) and $NH = 0.71G - 23.2$ for the Box (red line), where NH is the observed Alice surface brightness and G is the \galex\ surface brightness (Table \ref{tab:galexnh}). We have shown $1 \sigma$ error bars for the Box but not for the Stem to avoid clutter in the plot.}
    \label{fig:galexnh}
\end{figure}

\begin{table}[t]
\caption{Galex/Alice scale factor
\label{tab:galexnh}}
\begin{tabular}{lllc}
\hline\hline
 &r$^{a}$& Slope & Offset$^{b}$ \\
\hline
Stem & 0.742 & $ 0.82 \pm 0.14$ & $ -31.2 \pm 66.4$\\
Box   & 0.932 & $ 0.71 \pm 0.05$ & $ -23.2 \pm 24.1$\\
\hline
\hline
\multicolumn{4}{l}{$^a$ Correlation coefficient.}\\
\multicolumn{4}{l}{$^b$ \photu }\\
\end{tabular}
\end{table}

The Alice spectrograph was calibrated on the ground with the calibration updated through stellar observations during the mission. However, it is difficult to calibrate aperture-filling diffuse sources, especially for an aperture the size of the Box, and we have used diffuse \galex\ data \citep{Murthy2014apj} to cross-calibrate the Alice observations.
 We calculated the mean \galex\ FUV (1536~\AA) background in the Stem and the Box for each observation and compared it with the mean Alice surface brightness between 1400 -- 1700~\AA\ for each aperture. The errors in the Alice data were estimated from the mean of the errors in that range. We found an excellent correlation for both the Stem and the Box with the \galex\ diffuse background values (Fig. \ref{fig:galexnh}), but with slopes of $ 0.82 \pm 0.14$ and $ 0.71 \pm 0.05$, respectively (Table \ref{tab:galexnh}), reflecting both changes in the calibration over the lifetime of the mission and the difficulty in measuring the exact solid angle of the apertures. We have, therefore, rescaled the Alice spectra by these factors, assuming that a single scale factor applies over the entire Alice spectrum. There is evidence for a small offset ($\approx 23$ \photu\ at $1 \sigma$) between \galex\ and Alice,  perhaps due to two-photon emission arising in the Earth's atmosphere \citep{Kulkarni2022}.

\subsection{CUVB at $\lambda < 912$~\AA}

Because Alice sensitivity extends below the Lyman limit, we have used the spectral region from  600 -- 800~\AA\ to constrain 
the EUV radiation field, as measured from the outer solar system.  The mean surface brightness over all the observations is $3.2 \pm 3.0$ \photu\ in the Box and $-34.1 \pm\ 30.6$ \photu\ in the Stem. For radiation 
emerging from the field of the Alice box (4 deg$^2$ or $10^{-3}$ sr), 3.2 \photu\ corresponds to an 
EUV flux of 0.0032~photons~cm$^{-2}$~s$^{-1}$~\AA$^{-1}$.  \citet{Murthy_voy} also concluded that there was no emission shortward of the Lyman limit, but with less certainty given the relatively poor sensitivity of the \voyager\ ultraviolet spectrometer (UVS).

Currently, the best EUV flux limits come from studies by the {\it Extreme Ultraviolet Explorer} (EUVE) 
mission, which observed 54 local stars within 150~pc, mostly white dwarfs \citep{Dupuis1995, Vallerga1998}. The EUVE fluxes as seen from Earth are dominated by radiation from five stars:  
$\epsilon$~CMa ($d = 124$ pc), $\beta$~CMa (151~pc), G191-B2B (52.5~pc), HZ43A (60.3 pc), 
and Feige~24 (77.7 pc).  The NH-Alice limit  (0.0032~photons~cm$^{-2}$~s$^{-1}$~\AA$^{-1}$)
is well below the fluxes from these stars.  However, the local EUV radiation field is likely to be quite 
anisotropic, and the stellar  EUV fluxes have been attenuated by variable absorption from the local 
interstellar clouds.  In the wavelength range observed by EUVE, the brightest WD fluxes peak at 
250~\AA\ (owing to \HI\ and \HeI\ absorption) at 
0.15 -- 0.60 photons~cm$^{-2}$~s$^{-1}$~{\rm \AA}$^{-1}$, dropping to 
0.01 -- 0.04 photons~cm$^{-2}$~s$^{-1}$~\AA$^{-1}$ at 450 -- 600~\AA.  By comparison, 
$\epsilon$~CMa has photon flux at 600 -- 900~\AA\ ranging from 
0.6 -- 1.1 photons~cm$^{-2}$~s$^{-1}$~{\rm \AA}$^{-1}$.  This is consistent with a local EUV radiation field
that is highly directional and dominated by just a few stars, as noted by \citet{Murthy_sahnow2004} using observations at 1000 -- 1200~\AA\ from the FUSE spacecraft.

We can also compare the 3.2 \photu\ limit to the low-redshift metagalactic ionizing background flux,
which has been estimated to be
$J_0 = 1.3^{+0.8}_{-0.5}\times10^{-23}~{\rm erg~cm}^{-2}~{\rm s}^{-1}~{\rm Hz}^{-1}~{\rm sr}^{-1}$ \citep{Shull1999}
, which translates to 2.15 \photu\ in wavelength-distribution units.  The ionizing fluxes
decline at shorter wavelengths, and will be somewhat lower at 600~\AA. The metagalactic 
LyC photons are unlikely to propagate to the Galactic disk plane, owing to strong \HI\ absorption.

\subsection{Offsets at zero-reddening}

\begin{table}[t]
\caption{Correlation with E(B - V)
\label{tab:corr}}
\begin{tabular}{lllll}
\hline\hline
Quantity &r$^{a}$& Slope$^{b}$ & Offset$^{c}$ & $\chi^{2}$\\
\hline
\multicolumn{5}{c}{912 -- 1100~\AA}\\
\hline
STEM EBV & 0.596 & $2214 \pm 1694$ & $ 172 \pm  48$ & 0.15\\
BOX EBV & 0.807 & $2994 \pm  446$ & $ 221 \pm   11$ & 2.92\\\hline
\multicolumn{5}{c}{1400 -- 1700~\AA}\\
\hline
STEM EBV & 0.702 & $6554 \pm 4612$ & $ 257 \pm  130$ & 0.08\\
BOX EBV  & 0.909 & $6723 \pm  909$ & $ 264 \pm  24$ & 1.17\\

\hline
\multicolumn{5}{l}{$^a$ Correlation coefficient.}\\
\multicolumn{5}{l}{$^b$ \photu\ mag$^{-1}$}. \\
\multicolumn{5}{l}{$^c$ \photu.}\\
\end{tabular}
\end{table}

\citet{Akshaya2019} found that \galex\ FUV observations were tightly correlated with the Planck E(B~-~V) \citep{PlanckCollaboration2014} and with the 100 \micron\ emission from \citet{Schlegel1998}, with zero-offsets of 230 -- 290 \photu\ at both poles. They noted that these offsets were much greater than the $73 \pm 16$ \photu\ attributed to the integrated light of galaxies in the FUV \citep{Driver2016}, in agreement with earlier observations of the CUVB at the Galactic poles (Table \ref{tab:polar}). They attributed the excess emission to a new component of the diffuse background, but one not associated with interstellar dust \citep{Henry2015}. Similar conclusions were drawn by \citet{Hamden2013}, who presented an all-sky \galex\ map with a non-scattered isotropic component of diffuse FUV emission at a level of 300 \photu.  

\begin{figure}
    \includegraphics[width=3.5in]{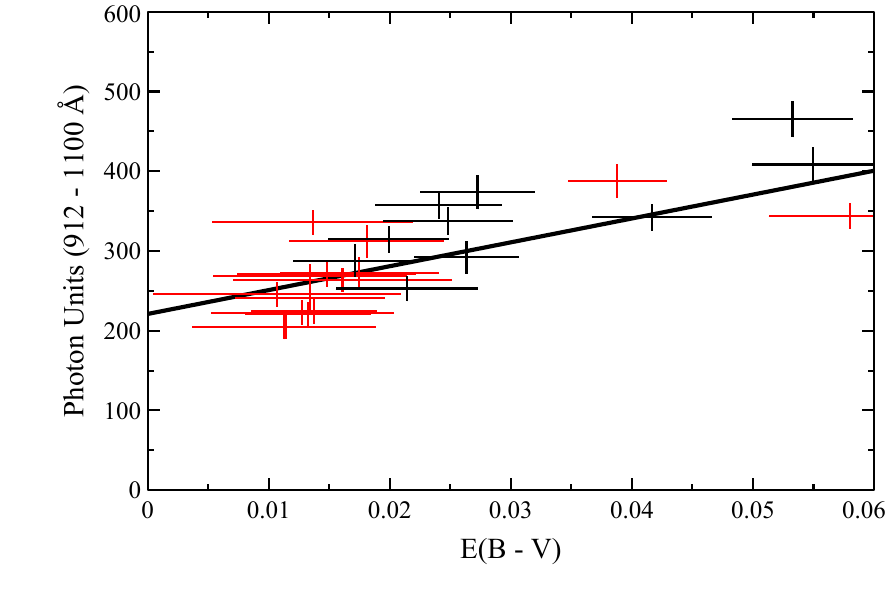}
    \caption{Mean surface brightness in the Box from 912 -- 1100~\AA\ as a function of the E(B~-~V) with $1 \sigma$ errors. Red points are near the SGP and black points are near the NGP. The line represents the best fit to the data with SB = 2994\ebv\ + 221 \photu\ (Table \ref{tab:nhbkgds}), , where SB is the observed surface brightness.}
    \label{fig:ebv1100}
\end{figure}

\begin{figure}
    \includegraphics[width=3.5in]{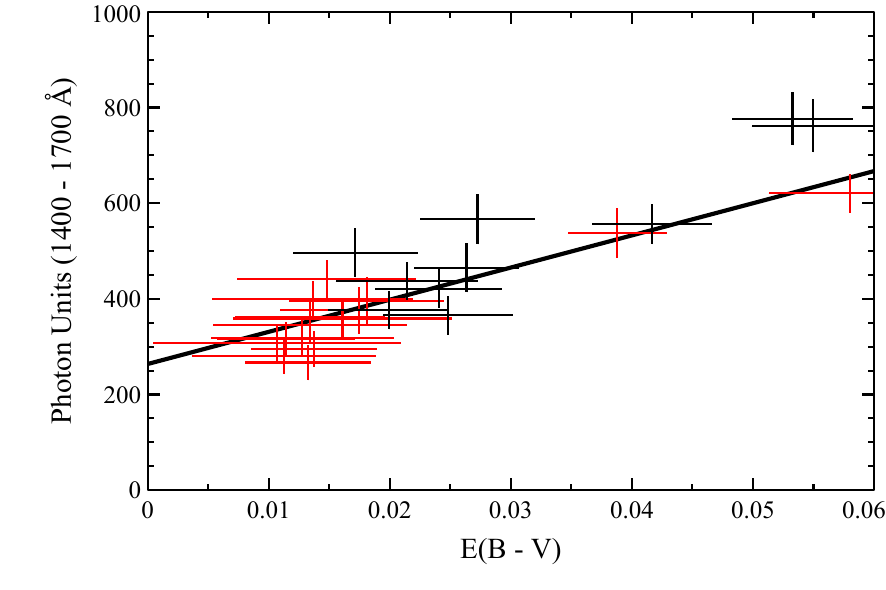}
    \caption{Mean surface brightness in the Box from 1400 -- 1700~\AA\ as a function of the E(B~-~V) with $1 \sigma$ errors. Red points are near the SGP and black points are near the NGP. The line represents the best fit to the data with SB = 6723\ebv\ + 264 \photu\ (Table \ref{tab:nhbkgds}), where SB is the observed surface brightness.}
    \label{fig:ebv1500}
\end{figure}

The Alice spectra offer an opportunity to test this correlation from a location near the edge of the Solar System, where we will only see emission from Galactic and extragalactic sources. We first divided the Alice spectra into two bands (912 -- 1100~\AA\ and 1400 -- 1700~\AA) and found the mean surface brightness in each band for the Stem and the Box, with the error assumed to be the mean error over the bandpass. Note that the line width of the Box (FWHM 172~\AA) meant that we had to integrate to 850~\AA\ to include all the signal from the short-wavelength band. The bandpass-averaged surface brightness is well correlated with the reddening in each aperture (Fig. \ref{fig:ebv1100} and \ref{fig:ebv1500}), where the reddening was obtained from the mean of the Planck E(B~-~V) over each aperture. The uncertainty in the E(B~-~V) was also taken from the Planck data and was approximately 0.005 mag in these fields.

\begin{figure*}
    \includegraphics{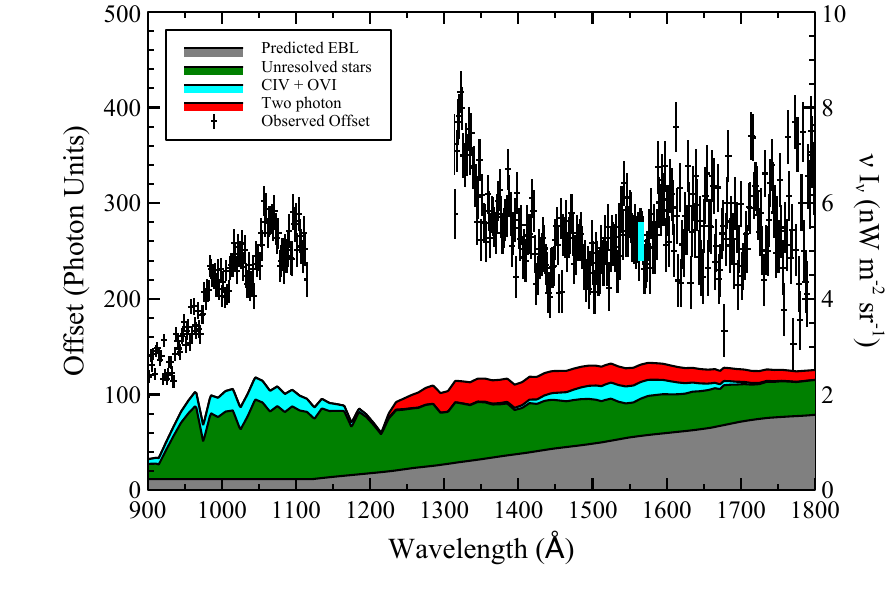}
    \caption{Offsets at zero reddening plotted as 1 $
    \sigma$ error bars for the Box. The discontinuity from 1100 -- 1350~\AA\ is because we have blanked out the section where uncertainties in the subtraction of the \lya\ template dominate the errors. Components of the diffuse radiation field are plotted in order from the bottom: EBL (filled grey) from \citet{Koushan2021}, unresolved stars (filled green), O VI and C IV line emission (cyan), and two-photon emission (red). The cyan bar represents the \galex\ offset found by \citet{Akshaya2018}. See text for discussion.}
    \label{fig:eblcomponents}
\end{figure*}

\begin{figure}
    \includegraphics[width=3.5in]{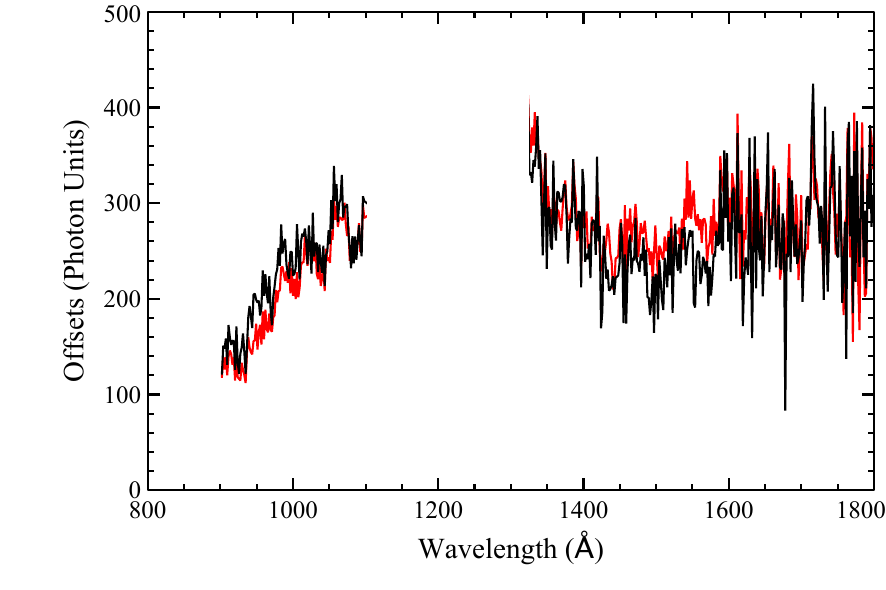}
    \caption{Offsets for the northern observations (black) and the southern observations (red).}
    \label{fig:eblpoles}
\end{figure}

We performed a least-squares fit between the UV surface brightening and the reddening, taking into account the uncertainties in both the Alice values and the E(B~-~V), and the resultant slopes and offsets are tabulated in Table \ref{tab:corr}. As expected, we find similar values from both the Stem and the Box but will focus on the Box results because of their much higher signal-to-noise ratio. The Box spectra are of sufficiently high quality that we can fit the offset, again using a least-squares fit, at each wavelength (Fig. \ref{fig:eblcomponents}). This is the first measurement of the zero-offset in the wavelength region between 912 -- 1100~\AA\ and is consistent with the $< 200$ \photu\ found through \voyager\ observations of the NGP from 912 -- 1100~\AA\ \citep{Holberg1986}, considering the relatively poor sensitivity of the \voyager\ UVS. The offsets are close to flat between 1400 -- 1800~\AA\ at a level of about 290 \photu, confirming previous observations made from Earth orbit (Table \ref{tab:polar}). Although it appears that the offsets decrease almost linearly from 1100 -- 912~\AA, we caution that this may be an artifact of the Box line width of 172~\AA. There is no significant difference between the offsets determined from observations in the southern hemisphere to those from the northern hemisphere (Fig. \ref{fig:eblpoles}). We will discuss the likely contributors below but note that known sources amount to less than half of the total observed offset.

The EBL in the UV is generally thought to be comprised primarily of the integrated light of galaxies \citep{Driver2016, Mattila_EBL2019}. This has been estimated
by several previous surveys, with values ranging from 73 to 195 \photu\ in the \galex\ FUV band. \citet{Gardner2000} used Space Telescope Imaging Spectrograph (STIS) observations of 
portions of the North and South Hubble Deep Fields and a parallel field near the North Deep Field
down to AB magnitude 29 -- 30 to find a range from 144 -- 195 \photu\ for the EBL, specifically quoting both
lower limits $\Phi_{\lambda} \geq 144^{+28}_{-19}$ \photu\ and 
upper limits $\Phi_{\lambda} \leq 195^{+59}_{-39}$ \photu.
A subsequent analysis of the Hubble Ultraviolet Ultra-Deep Field \citep{Driver2016} in the 
FUV (1530~\AA\ band) quoted a value of $1.45 \pm 0.27$ nW~m$^{-2}$~sr$^{-1}$,
corresponding to $73 \pm 14$ \photu.  Their analysis included number counts of galaxies from 
\galex\ extending to AB magnitude 23.8, yielding $34 \pm 5$ \photu\ \citep{Xu2005};  their
survey extended 5 magnitudes deeper, to AB = 29 -- 30, using the solar-blind channel of
the HST Advanced Camera for Surveys.  Both HST surveys showed that galaxy counts 
increase at AB $>23$ mag, flattening from AB = 26 to 30. We have adopted the value of $73 \pm 14$ \photu\ from \citet{Driver2016} for the contribution to the FUV background from galaxy counts at high Galactic latitudes and used the spectrum from \citet{Koushan2021}.

Only hot (O and B) stars can contribute significantly in the Alice spectral range, particularly in the 900 -- 1100~\AA\ band. We checked each field using Astroquery \citep{Ginsburg2019} for the presence of any O, B, or A stars near the Box in the Simbad database \citep{Wenger2000}. Only four fields included A stars, with no O or B stars, with an effective contribution of about 30 \photu\ at 1500~\AA\ and nothing at 1000~\AA. These stars will be included in the \galex\ point source catalog \citep{Bianchi2018}, which is complete to an AB magnitude of 19.9 in the FUV. Stars in the catalog contribute a mean of $26.7 \pm 10.5$ \photu\ in the \galex\ FUV band (1350 -- 1800~\AA). We estimated the contribution of fainter stars to $30^{th}$ mag using TRILEGAL \citep{Girardi2005} and found that the expected contribution was equivalent to one B1 star per square degree with an AB magnitude of 17.7, equivalent to a diffuse signal of $19.1 \pm 2$  \photu\ and $13.6 \pm 2$ \photu\ at 1000~\AA\ and 1500~\AA, respectively. We assumed that the stars from the \citet{Bianchi2018} catalog have a similar spectrum and find a total contribution of $56.2 \pm 11$ \photu\ at 1000~\AA\ and $40.3 \pm 11$ \photu\ at 1500~\AA\ from resolved and unresolved stars.

O VI (1032/1038~\AA) emission from the Galactic halo will contribute $4450 \pm 950$ ph cm$^{-2}$ s$^{-2}$ sr$^{-1}$ \citep{Dixon2001, Shelton2001, Jo2019} and C IV (1548/1550~\AA) will contribute approximately $5000 \pm 800$ ph cm$^{-2}$ s$^{-2}$ sr$^{-1}$ \citep{Martin1990_h2, Jo2019}. Because of the 172~\AA\ FWHM of the Box, these lines are spread over the entire spectral range and correspond to a contribution of about 20 \photu\ in each band. Finally, two-photon emission from the warm ISM will contribute an additional 20 \photu\ \citep{Kulkarni2022}. There may also be small contributions from H$_{2}$ fluorescence in the Lyman bands between 1330 and 1620~\AA\, if some fields contain trace amounts of H$_{2}$ in the cirrus clouds \citep{Gillmon2006, Jo2017}.

We have plotted the different components discussed above in Figure \ref{fig:eblcomponents}, finding that the observed radiation is at least double the sum of all the contributors over the entire spectral range. We have, as yet, no explanation for the excess radiation. Based on spatial cross-correlation with redshift surveys of extragalactic objects, \citet{Chiang2019} identified a monopole contribution of 90 (+28,-16) Jy sr$^{-1}$ (approximately 91 \photu) of FUV background associated with extragalactic objects.  They argue that an extragalactic origin to the unaccounted foreground can be ruled out by their clustering analysis.

\section{Implications for Future Missions}

Future experiments relevant to the FUV background include NASA's recently selected mission UVEX (UltraViolet EXplorer) to survey ultraviolet light across the entire sky, providing insight into how galaxies and stars evolve. UVEX is targeted to launch in 2030 as NASA’s next Astrophysics Medium-Class Explorer mission \url{(https://www.uvex.caltech.edu)}. In addition to its capability to quickly point toward sources of ultraviolet light in the universe, UVEX will conduct a highly sensitive all-sky survey at both FUV and NUV wavelengths, with a sensitivity 50 times better than \GALEX.  UVEX will have a 
wide-field ($3.5^{\circ} \times 3.5^{\circ}$)  two-band ultraviolet imager covering the 
FUV (1390--1900~\AA) and NUV (2030--2700~\AA) with $2''$ point spread function.  Ultraviolet
spectroscopy will be obtained with a two-degree long multi-width slit spectrometer covering 
wavelengths from 1150~\AA\ to 2650~\AA.  In its wide elliptical orbit around the Earth 
($17-59~R_{\rm E}$) and synoptic observations, UVEX should be above the Earth's exospheric
contributions (Kulkarni 2022) and able to monitor (and subtract out) most of the solar contributions 
to the FUV background. As with the cosmic X-ray background, it is likely that UVEX will uncover a more complex picture
of the spatial and spectral signatures of the CUVB.  
  
\section{Summary}

\begin{table}[t]
\caption{Components of Offsets
\label{tab:summary}}
\begin{tabular}{lll}
\hline\hline
Source & Offset$^{a}$ & Ref.\\
\hline
\multicolumn{3}{c}{912 -- 1100~\AA}\\
\hline
Observed  & $ 221 \pm  11$ & This work\\
EBL & $< 10$ & \citet{Koushan2021}\\
Stars & $56 \pm 11$ & \citet{Girardi2005}\\
& & \citet{Bianchi2018}\\
O VI & $22 \pm 5$ & \citet{Shelton2001}\\
\hline
Excess & $133 \pm 17$ \\
\hline
\multicolumn{3}{c}{1400 -- 1700~\AA}\\
\hline
Observed  & $ 264 \pm 24$ & This work\\
EBL & $73 \pm 16$ & \citet{Driver2016}\\
Stars & $40 \pm 11$ & \citet{Girardi2005}\\
& & \citet{Bianchi2018}\\
CIV & $25 \pm 4$ & \citet{Martin1990_lines}\\
Two-photon & $23 \pm 3$ & \citet{Kulkarni2022}\\
\hline
Excess & $103 \pm 31$\\
\hline
\multicolumn{3}{l}{$^a$ \photu.}\\
\end{tabular}
\end{table}

We have observed 25 targets near the Galactic poles with the Alice spectrograph on the New Horizons spacecraft. These target fields were complementary to LORRI observations of the COB \citep{Postman2024}. The CUVB is linearly correlated with the interstellar reddening with offsets of  $221 \pm 11$ \photu\ at 1000~\AA\ and $264 \pm 24$ at 1500~\AA, consistent with earlier observations of the offsets at the Galactic Poles and significantly greater than the sum of the known sources (Table \ref{tab:summary}). The excess surface brightness over all the identified sources is $133 \pm 17$ \photu\ at 1000~\AA\ and $103 \pm 31$ \photu\ at 1500~\AA.

There is no background radiation detected shortward of the \HI\ Lyman limit in any of the high-latitude fields, at a level of $3.2 \pm 3.0$ \photu\ between 600 -- 800~\AA.  This corresponds to an EUV flux of 0.003 photons~cm$^{-2}$~s$^{-1}$~\AA$^{-1}$, suggesting that the EUV emission from the bright stellar sources seen by EUVE  (Dupuis \etal\ 1995; Vallerga 1998) is highly localized.

Importantly, we have found that there is, at most, an offset of $-23 \pm 24$ \photu\ between the Alice observations and \galex\ data at 1500~\AA, that is that the contributions to the diffuse background are close to zero from the Earth's atmosphere. Hence, observations of the FUV (1300 -- 1800~\AA) may very well be made from spacecraft in low Earth orbit, such as \galex\ or the future UVEX mission (see below). This is not true for observations shortward of 1200~\AA, where scattering from the intense \lya\ line will dominate the signal unless we observe from the outer Solar System.

\section{Acknowledgments}
\begin{acknowledgments}
We thank NASA for funding and continued support of the New Horizons mission, which were required to obtain the present observations. No New Horizons NASA funds were used for the analysis and writing of this paper. The data presented were obtained during the second Kuiper Extended Mission of New Horizons.
We thank Kwang-Il Seon, Jonathan Gardner, Harry Teplitz, and Shri Kulkarni for useful discussions. Some of the data presented in this paper were obtained from the Mikulski Archive for Space Telescopes (MAST). STScI is operated by the Association of Universities for Research in Astronomy, Inc., under NASA contract NAS5-26555. Support for MAST for non-HST data is provided by the NASA Office of Space Science via grant NNX13AC07G and by other grants and contracts. This research has made use of the SIMBAD database, CDS, Strasbourg Astronomical Observatory, France.
All costs associated with the publication of this paper were borne by NASA’s Maryland Space Grant Consortium.
\end{acknowledgments}

\vspace{5mm}
\facilities{New Horizons Alice}

\software{GnuDataLanguage \citep{GDL2010, GDL2011, GDL2022}, Fawlty Language \textit{http://www.flxpert.hu/fl/}, TRILEGAL \citep{Girardi2005}}

\bibliography{murthy}{}
\bibliographystyle{aasjournal}

\end{document}